\documentclass[a4paper,11pt]{article}
\usepackage{pos}
\usepackage{wrapfig}
\usepackage{graphicx}
\usepackage{caption}
\usepackage{verbatim}
\usepackage{slashed}
\usepackage{soul}
\usepackage{xcolor}
\title{Meson mixing effects on the speed of sound in isospin-imbalanced matter}

\ShortTitle{Meson mixing effects on the speed of sound in isospin-imbalanced matter}

\author*[a]{Alejandro Ayala}
\author[b,c]{Bruno S. Lopes}
\author[b,c]{Ricardo L. S. Farias}
\author[a]{Luis C. Parra}

\affiliation[a]{Instituto de Ciencias Nucleares, Universidad Nacional Autónoma de México,\\
  Circuito exterior S/N Ciudad Universitaria, CDMX, 04510, México}

\affiliation[b]{Departamento de Física, Universidade Federal de Santa Maria,\\
RS 97105-900, Rio Grande do Sul, Brazil}

\affiliation[c]{Center for Nuclear Research, Department of Physics, Kent State University,\\
Kent, OH 44242, USA}

\emailAdd{ayala@nucleares.unam.mx}
\emailAdd{bruno.lopes@acad.ufsm.br}
\emailAdd{ricardo.farias@ufsm.br}
\emailAdd{luis.parra@correo.nucleares.unam.mx}

\abstract{We explore isospin-imbalanced strongly interacting matter within the two-flavor Linear Sigma Model with quarks, an effective model for low-energy QCD. At one-loop order, including quark, pion, and sigma fluctuations while respecting chiral symmetry, we find that the formation of an isospin condensate necessarily gives rise to a Goldstone mode. This mode enforces a nontrivial relation between the chiral and isospin condensates through the mixing of charged pions and the sigma field in the condensed phase. From the resulting thermodynamic potential, we compute the speed of sound and observe a pronounced peak as a function of the isospin chemical potential. Although the peak of the speed of sound may be described at tree-level and including only quarks in the analysis, meson dynamics introduces further constraints that influence the position and width of the peak which making it to align well with lattice QCD simulations. Therefore we identify that the shape and position of the peak is a consequence of the Goldstone mode dynamics and of the associated charged pion–sigma mixing.}

\FullConference{53rd International Symposium on Multiparticle Dynamics\\ September, 9th. 2025.
}

\tableofcontents

\begin{document}
\maketitle

\section{Introduction}

With the advent of the first particle accelerators, the meaning of ``high energy physics'' expanded beyond merely high temperatures. It first grew to encompass high densities and, more recently, any configuration where energy is stored in the fundamental degrees of freedom of particles such as in strong magnetic fields or, as in this work, isospin-dense matter. At energy scales that exceed nuclear densities and overcome electro-weak interactions between nucleons, the dominant force arises from the interactions between subatomic particles: quark--quark, quark--gluon, and gluon--gluon. When the strongest force in the universe takes center stage, the theoretical description becomes highly complex and often computationally intractable. For that reason, one may instead focus on the most relevant degrees of freedom for the specific conditions of interest, simplifying and clarifying the overall picture. In this work, we establish the range of validity for the Linear Sigma Model with quarks (LSMq) in describing cold, strongly interacting matter. This allows us to investigate the origin of an intriguing phenomenon: a peak in the speed of sound of dense matter, inferred from meta-analyses of neutron star mass-radius observations as in Ref.\cite{Kojo2021} and yet to be fully understood theoretically.

\subsection{Isospin imbalanced strongly interacting matter}

In extreme astrophysical environments such as neutron star mergers and the cores of massive neutron stars, or speculatively in some high energy collisions, nuclear matter exists under conditions of significant isospin asymmetry. This occurs when the densities of neutrons ($n$) and protons ($p$) are substantially unpaired, quantified by the isospin chemical potential $\mu_I = - (\mu_n - \mu_p)/2$, or similarly in terms of the light quark chemical potentials ($u,d$) as $\mu_I = (\mu_u - \mu_d)/2$. Under such imbalance, the physics of the strong interaction exhibits phenomena distinct from symmetric matter. In particular, for sufficiently large $\mu_I$, greater than the pion mass $m_\pi$, charged pion condensation may occur, fundamentally restructuring the ground state with the emergence of a Goldstone boson.

This system is governed by Quantum Chromodynamics (QCD), and the theoretical challenge lies in describing this regime from first principles A highly successful approach in this context is given by lattice QCD (LQCD), which can perform simulations at finite isospin density but suffers from the sign problem at nonzero baryon chemical potential. Perturbative methods, while reliable within its region of validity, are not applicable at the intermediate to high densities relevant to neutron star phenomenology. These limitations make effective models like the LSMq or Nambu--Jona-Lasinio model (NJL), both chiral models, particularly valuable tools for mapping out the equation of state (EoS) and transport properties in this theoretically challenging region of the QCD phase diagram.

\subsection{QCD phase diagram}

The phase structure of QCD as a function of temperature $T$ and baryon chemical potential $\mu_B$ represents one of the most fundamental, theoretically and experimentally rich maps of matter in the universe. While the high-temperature, low-density regime of quark-gluon plasma is accessible via heavy-ion collisions and LQCD, the high-density, low-temperature region relevant for neutron star interiors remains a largely unexplored territory. A particularly significant landmark within this diagram is the hypothesized Critical End Point (CEP) of QCD, at which the change from the hadronic phase to the quark-gluon plasma goes from a smooth crossover to a first-order phase transition (precisely at the CEP, the transition is of second-order). Locating the CEP is one of the major goals in the field. However, as pointed out in Ref.~\cite{Mannarelli2019}, current theoretical and experimental studies have yet to probe its expected region in the $\mu_B$-$T$ plane. This inaccessibility motivates substantial efforts to develop effective models capable of describing strongly interacting matter in these extreme regions, or in analogous regimes where key physical features (like critical behavior) may be studied.

\subsubsection*{Multidimensional QCD phase diagram}

The full QCD phase space is inherently multidimensional. Beyond temperature and baryon chemical potential, additional parameters control the physical system as seen in Table~\ref{tab:qcd_parameters}.
\begin{table}[ht]
    \centering
    \begin{tabular}{|p{4cm}|p{10cm}|}
        \hline
        \textbf{Parameter} & \textbf{Physical Role} \\
        \hline
        Baryon chemical potential $\mu_B$ & Controls net baryon density \\
        \hline
        Isospin chemical potential $\mu_I$ & Controls neutron-proton asymmetry \\
        \hline
        Chiral chemical potential $\mu_5$ & Related to chirality imbalance \\
        \hline
        Axial-isospin chemical potential $\mu_{I5}$ & Controls axial charge in isospin space \\
        \hline
        Magnetic field $B$ & Breaks $O(3)$ symmetry, quantizing charged particle motion into discrete Landau levels \\
        \hline
        Rotation/vorticity $\omega$ & Selects a preferred frame of reference, introducing a coupling to the total angular momentum $\vec{\omega} \cdot \vec{J}$ \\
        \hline
        Strangeness $\mu_S$ & Controls net strangeness in the system \\
        \hline
    \end{tabular}
    \caption{Control parameters in the multidimensional QCD phase diagram.}
    \label{tab:qcd_parameters}
\end{table}
These additional dimensions dramatically enrich the phase structure, introducing new critical lines and exotic phases such as pion/kaon condensation and magnetic catalysis. Additionally, the use of dual symmetries between different phases offers a pathway to explore previously inaccessible regions.

\subsubsection*{Dual transformations and phase interrelations}

In the presence of multiple chemical potentials ($\mu_B$, $\mu_I$, $\mu_5$, $\mu_{I5}$), massless NJL and QCD Lagrangians exhibit invariance under certain dual transformations~\cite{Zhokhov2025}. These transformations interchange order parameters and map between different phases:

\begin{itemize}
    \item \textbf{Three-color QCD:} Chiral Symmetry Breaking (CSB) $\Leftrightarrow$ Charged Pion Condensation (CPC)
    \item \textbf{Two-color NJL:} Additional dualities: CSB $\Leftrightarrow$ CPC $\Leftrightarrow$ Baryon Superfluid (BSF)
\end{itemize}

These dualities create symmetric structures in the combined space of chemical potentials. Knowledge of one phase for a given set of chemical potentials directly informs the phase structure in dual regions, effectively reducing the computational work by mapping solutions rather than recomputing everywhere.

\subsubsection*{The sign problem}

At finite baryon density ($\mu_B > 0$), the QCD fermion determinant becomes complex, making standard LQCD Monte Carlo methods inapplicable and defining what is known as the ``sign problem". While it can be circumvented through expansions of the pressure for up to $\mu_B / T \lesssim 2-3$, the entire neutron star core region remains inaccessible due to its very high density and comparably small temperature. This theoretical impasse at finite density is precisely what makes the neutron star EoS a fundamental problem in nuclear astrophysics. The multidimensional nature of the QCD phase diagram, combined with the sign problem, justifies the necessity of developing sophisticated effective models that can capture both the rich phase structure and the dual symmetries present in the theory.

\subsubsection*{Effective models}

In the absence of first principles access, effective field theories and phenomenological models become essential tools. These models capture the relevant symmetries of QCD, incorporate known low-energy degrees of freedom (pions, nucleons, quarks), are tractable at finite density and temperature and can be constrained by LQCD where available ($T > 0$, $\mu_B \sim 0$).

\subsection{Isospin chemical potential}\label{sec:isospin_mu}

In the presence of two active light flavors, an isospin chemical potential $\mu_I$ can be introduced to study matter with an imbalance between up and down quark number densities. The individual quark chemical potentials are then expressed in terms of the baryon chemical potential $\mu_B$ and $\mu_I$ as $
\mu_u = \frac{\mu_B}{3} + \frac{\mu_I}{2}$ and $
\mu_d = \frac{\mu_B}{3} - \frac{\mu_I}{2}$.

In the special case of zero baryon density ($\mu_B = 0$), the quark chemical potentials reduce to $\mu_u = \mu_I/2$ and $\mu_d = -\mu_I/2$, which corresponds to a system with equal magnitudes and opposite signs for the chemical potentials of up and down quarks. This setup is particularly relevant for studying pion condensation, as the effective pion chemical potential is given by $\mu_{\pi_+} = \mu_u - \mu_d = \mu_I$. The isospin number density, which measures the net imbalance between up and down quarks, is defined as $n_I = n_u - n_d$.

\subsection{Chirality and isospin}

The chiral and isospin condensates are dynamically affected by the isospin chemical potential. At $\mu_I = 0$ chiral symmetry is spontaneously broken and the chiral condensate takes a nonzero value. As $\mu_I$ increases, the system remains in the chirally broken phase until a critical value $\mu_I \approx m_\pi$ is reached~\cite{Andersen2022}.

Beyond this threshold, the system enters the pion-condensed phase, signaling the onset of Bose-Einstein condensation (BEC). In this regime, the chiral condensate decreases monotonically. At sufficiently large $\mu_I$, the condensates undergo a smooth crossover into a phase reminiscent of Bardeen-Cooper-Schrieffer (BCS) pairing among quarks. After the transition, the residual $U(1)_{I_3}$ is broken, corresponding to a condensation of one of the charged pions and thus to the development of a Goldstone boson mode.

\subsection{Speed of sound}

Recent studies have investigated the behavior of the speed of sound squared \(c_s^2\) as a function of the isospin density \(n_I/n_0\) (with $n_0 \sim 0.16$ fm$^{-3}$ denoting the nuclear saturation density). Ref.~\cite{Chiba2024}, for example, highlights the importance of quantum corrections in accurately describing pion condensation within the framework of the LSMq (referred to as quark-meson model in this case). When quark loops are included, treating mesons as composite rather than elementary fields, the isospin condensate saturates to a constant value at large \(\mu_I\). This saturation gives rise to an EoS of the form $P(\mu_I) \sim a_0 \mu_I^4 + a_2 \mu_I^2$, which drives the speed of sound squared \(c_s^2 = \partial P / \partial \epsilon\) toward qualitative agreement with lattice QCD simulations. This demonstrates that quantum corrections are essential for capturing the correct physics in the high-density regime. With some caveats, such as the use of a comparatively large coupling constant and the fact that meson loops are not explicitly included, this study provides important guidance for subsequent work, underscoring the role of quantum fluctuations in the LSMq to describe the isospin-dense region of the QCD phase diagram.

\section{Linear Sigma Model with quarks}

The LSMq provides an effective model framework to describe the low energy dynamics of QCD, incorporating both mesonic degrees of freedom and quark fields. It is particularly well suited for studying chiral symmetry breaking and pion condensation in the presence of an isospin chemical potential.

\subsection{Lagrangian}

The Lagrangian for the two-flavor LSMq is given by
\begin{equation}
\mathcal{L} = \frac{1}{2}(\partial_\mu \sigma)^2 + \frac{1}{2}(\partial_\mu \vec{\pi})^2 + \frac{a^2}{2}(\sigma^2 + \vec{\pi}^2) - \frac{\lambda}{4}(\sigma^2 + \vec{\pi}^2)^2 + i\bar{\psi}\slashed{\partial} \psi - ig\bar{\psi}\gamma^5 \vec{\tau} \cdot \vec{\pi}\psi - g\bar{\psi}\psi\sigma,
\end{equation}
where \(\vec{\tau} = (\tau_1, \tau_2, \tau_3)\) are the Pauli matrices. The quark fields \(\psi\) are \(SU(2)_{L,R}\) doublets,
$\psi_{L,R} = (u, d )^T_{L,R}$,
\(\sigma\) is a real scalar field and \(\vec{\pi} = (\pi_1, \pi_2, \pi_3)^T\) is a triplet of real scalar fields. The parameters \(a^2\), \(\lambda\) and \(g\) are real and positive definite.

The field \(\pi_3\) corresponds to the neutral pion, while the charged pions are represented by the combinations
\begin{equation}
\pi_- = \frac{1}{\sqrt{2}}(\pi_1 + i\pi_2), \quad \pi_+ = \frac{1}{\sqrt{2}}(\pi_1 - i\pi_2).
\end{equation}
In terms of the charged and neutral pion degrees of freedom, the Lagrangian can be written as
\begin{align}
\mathcal{L} &= \frac{1}{2} [(\partial_\mu \sigma)^2 + (\partial_\mu \pi_0)^2] + \partial_\mu \pi_- \partial^\mu \pi_+ + \frac{a^2}{2} (\sigma^2 + \pi_0^2) + a^2 \pi_- \pi_+ \nonumber \\
&\quad - \frac{\lambda}{4} (\sigma^4 + 4\sigma^2 \pi_- \pi_+ + 2\sigma^2 \pi_0^2 + 4\pi_-^2 \pi_+^2 + 4\pi_- \pi_+ \pi_0^2 + \pi_0^4) \nonumber \\
&\quad + i\bar{\psi} \slashed{\partial} \psi - g \bar{\psi} \psi \sigma - ig \bar{\psi} \gamma^5 (\tau_+ \pi_+ + \tau_- \pi_- + \tau_3 \pi_0) \psi,
\end{align}
where we have introduced the charged isospin matrices
$\tau_+ = \frac{1}{\sqrt{2}} (\tau_1 + i\tau_2)$,  $\tau_- = \frac{1}{\sqrt{2}} (\tau_1 - i\tau_2)$,
which define the charged pion basis. The neutral pion, by contrast, couples to $\tau_3$, which remains unchanged in this basis.

\subsection{Symmetries and their breaking}

The Lagrangian of the LSMq is constructed to reflect the fundamental symmetries of low energy QCD. In the chiral limit, it exhibits a global \(SU(N_c)\) color symmetry, a \(U(1)_B\) baryon number symmetry, and most importantly, a \(SU(2)_L \times SU(2)_R\) chiral symmetry. This chiral symmetry plays a central role in the low-energy dynamics of strong interactions, governing the relationship between quarks and mesons.

However, in nature, these symmetries are not all realized in the ground state of the system. They can be broken in different ways, each giving rise to distinct physical phenomena. In this work, we consider two such mechanisms: explicit chiral symmetry breaking, which accounts for the finite mass of the pions in the vacuum, and symmetry breaking induced by a finite isospin chemical potential, which drives the system into a pion-condensed phase.

\subsubsection*{Explicit symmetry breaking: finite pion masses}

In the real world, chiral symmetry is not exact: it is explicitly broken by the small but nonzero quark masses. This explicit breaking is responsible for the finite mass of the pions, which would otherwise be exact Goldstone bosons. To incorporate this into the model, we add a linear term in the sigma field to the Lagrangian, $\mathcal{L} \to \mathcal{L} + h \, \sigma$, which tilts the `Mexican hat' potential and lifts the degeneracy of the vacuum. The constant \(h\) is fixed by matching the model to the physical pion mass in the vacuum,
\begin{equation}
h = m_\pi^2 f_\pi,
\end{equation}
where \(f_\pi\) is the pion decay constant. This explicit breaking selects a unique ground state in which the sigma field acquires a non-vanishing vacuum expectation value, $\langle \sigma \rangle = v$, while the pion fields remain zero. This vacuum expectation value spontaneously breaks the remaining chiral symmetry and provides the quarks with a constituent mass \(m_f = g v\).

\subsubsection*{Isospin chemical potential: breaking isospin symmetry}

While the explicit chiral symmetry breaking term gives the pions their mass, it does not distinguish between the different isospin components, keeping pions degenerate. To explore phases with an isospin imbalance, we need to explicitly break the isospin symmetry by introducing a finite isospin chemical potential \(\mu_I\), as seen in Table~\ref{tab:qcd_parameters}. It couples to the conserved isospin charge, effectively favoring states with a net isospin density. In practice, this is implemented by modifying the derivatives in the Lagrangian to include a background field,
\begin{equation}
\partial_\mu \to D_\mu = \partial_\mu + i \mu_I \delta^0_\mu, \quad \partial^\mu \to D^\mu = \partial^\mu - i \mu_I \delta^\mu_0.
\end{equation}
This modification shifts the energies of particles and antiparticles according to their isospin quantum numbers. For the quarks, this translates into different chemical potentials for up and down flavors, $\mu_u = \frac{\mu_I}{2}$, $\mu_d = -\frac{\mu_I}{2}$, in the limit of zero baryon density. At finite isospin chemical potential, the isospin $SU(2)_I$ symmetry is explicitly broken down to the subgroup $U(1)_{I_3}$. As $\mu_I$ increases, it eventually becomes favorable for the system to develop a charged pion condensate, giving rise to a rich phase structure characterized by the interplay between the chiral condensate $v$ and the pion condensate $\Delta$ (for a detailed analysis in the NJL model, see Ref.~\cite{Mu2010zz}).

\subsubsection*{Pion condensation}

In the pseudoscalar channels, further simplifications can be obtained using the ansatz $\langle \pi^0 \rangle=\langle \bar{\psi} i \gamma_5 \tau_3 \psi \rangle = 0$, $ \langle \pi^- \rangle=\langle \bar{u} i \gamma_5 d \rangle = \langle \bar{d} i \gamma_5 u \rangle^* \neq 0$. This structure further breaks the residual \(U(1)_{I_3}\) symmetry, and corresponds to a BEC of one of the charged pions. The charged pion fields can be referred from their condensates by the shift
$\pi_+ \to \pi_+ + \frac{\Delta}{\sqrt{2}} e^{i\theta}$, $\quad \pi_- \to \pi_- + \frac{\Delta}{\sqrt{2}} e^{-i\theta}$,
where the phase factor \(\theta\) indicates the direction of the \(U(1)_{I_3}\) symmetry breaking. For definiteness, we take \(\theta = \pi\). However, all physical observables are independent of $\theta$, reflecting the Goldstone nature of the mode and ensuring the uniqueness of the physical solutions.

\subsubsection*{Masses in the condensed phase}

The shift in the sigma field and the presence of the pion condensate produce masses for the fermions and neutral bosons given by
\begin{align}
m_f &= gv, \\
m_{\pi_0}^2 &= \lambda \left( v^2+ \Delta^2\right) - a^2, \\
m_\sigma^2 &= \lambda \left(3 v^2+ \Delta^2\right) - a^2.
\end{align}

\subsection{Effective potential}

Since the Lagrangian contains anharmonic interaction terms beyond quadratic order, we approximate the full effective potential by including the first quantum correction. At one-loop order, the effective potential is given by $V_{\text{eff}} = V_{\text{tree}} + V^{(1)}$, where \(V^{(1)} = V^{(1)}_f + V^{(1)}_b\) receives separate contributions from the fermion and boson sectors. In the condensed phase, the tree-level potential takes the form
\begin{equation}
V_{\text{tree}} = -\frac{a^2}{2} \left( v^2 + \Delta^2 \right) + \frac{\lambda}{4} \left( v^2 + \Delta^2 \right)^2 - \frac{1}{2} \mu_I^2 \Delta^2 - hv.
\end{equation}
At this level, $v$ and $\Delta$ are free parameters, and they are obtained from the gap equation. While $v$ is determined by solving the gap equation derived from the full one-loop effective potential,
\begin{equation}
\left.\frac{dV_{\text{eff}}^{(1)}}{dv}\right|_{v,\Delta} = 0 \, ,
\end{equation}
the isospin condensate $\Delta$ is constrained by the requirement of a massless Goldstone mode associated with the broken $U(1)_{I_3}$ symmetry. This condition, which we analyze in detail in Sec.~\ref{sec:Goldstone}, yields specific solutions for $\Delta$ as a function of $v$ and $\mu_I$.
%
\subsubsection*{Fermion contribution to the effective potential}

The fermion contribution to the one-loop effective potential is given by
\begin{equation}
\sum_{f=u,d} V_f^1 = -2N_c \int \frac{d^3 k}{(2\pi)^3} \left( E_\Delta^u + E_\Delta^d \right),
\end{equation}
where the quasiparticle energies for up and down quarks in the condensed phase are
\begin{align}
E_\Delta^u &= \left[\left( \sqrt{k^2 + m_f^2} + \frac{\mu_I}{2} \right)^2 + g^2 \Delta^2 \right]^{1/2}, \\
E_\Delta^d &= \left[ \left( \sqrt{k^2 + m_f^2} - \frac{\mu_I}{2} \right)^2 + g^2 \Delta^2 \right]^{1/2}.
\end{align}

\subsubsection*{Boson contribution}

The boson contribution to the one-loop effective potential is more involved due to the mixing between the \(\sigma\) field and the charged pions in the condensed phase. It is given by the functional determinant

\begin{equation}
V_b^1 = \frac{i}{2} \int \frac{d^4 k}{(2\pi)^4} \ln \left[ \det \left( D_b^{-1} \right) \right],
\end{equation}
where the inverse boson propagator in the basis \((\sigma, \pi_+, \pi_-, \pi_0)\) takes the form
\begin{equation}
D_b^{-1} =
\begin{pmatrix}
K^2 - m_\sigma^2 & -\sqrt{2}\lambda v\Delta e^{-i\theta} & -\sqrt{2}\lambda v\Delta e^{i\theta} & 0 \\
-\sqrt{2}\lambda v\Delta e^{i\theta} & K^2 - m_{ch}^2 + \mu_I^2 + 2\mu_I k_0 & -\lambda\Delta^2 e^{2i\theta} & 0 \\
-\sqrt{2}\lambda v\Delta e^{-i\theta} & -\lambda\Delta^2 e^{-2i\theta} & K^2 - m_{ch}^2 + \mu_I^2 - 2\mu_I k_0 & 0 \\
0 & 0 & 0 & K^2 - m_{\pi_0}^2
\end{pmatrix}.
\end{equation}
Here \(K^2 = k_0^2 - \mathbf{k}^2\), and it is useful to define the mass-like combination
\begin{align}
m_{ch}^2 &= \lambda (v^2 + 2\Delta^2) - a^2,
\end{align}
which is not the fully dynamical charged pion mass. As evident from the non-diagonal entries, the \(\sigma\) field and the charged pions mix in the condensed phase, which must be properly accounted for when computing quantum corrections.

\subsection{Goldstone Mode Condition}
\label{sec:Goldstone}
The breaking of the \(U(1)_{I_3}\) global symmetry comes together with the development of a Goldstone boson. To find the restrictions imposed on \(v\) and \(\Delta\) by the appearance of a massless mode, we examine the limit \(K^\mu \to 0\) of the inverse boson propagator and require that its determinant vanishes, $\det D_b^{-1} = 0$. This is equivalent to keeping only the product of the masses in the calculation of this determinant. Explicitly, the determinant factorizes as
\begin{equation}
\det D_b^{-1} = m_{\pi^0}^2 \, m_\sigma^2 \, (m_{\pi^0}^2 - \mu_I^2) \left( m_{\pi^0}^2 + 2\Delta^2 \lambda \frac{m_{\pi^0}^2}{m_\sigma^2} - \mu_I^2 \right) = 0.
\end{equation}

There are three possible solutions for \(\Delta\) arising from this condition, out of which only two are real on the whole \(\mu_I \geq m_\pi\) domain:
\begin{align}
\Delta_1 &= \sqrt{\frac{\mu_I^2 - 2(3\lambda v^2 - 2a^2) + \sqrt{4a^4 + 4\mu_I^2(6\lambda v^2 - a^2) + \mu_I^4}}{6\lambda}}, \\
\Delta_2 &= \sqrt{\frac{\mu_I^2 - (\lambda v^2 - a^2)}{\lambda}}.
\end{align}
The solution \(\Delta_2\) corresponds to the condition for the existence of a massless mode at tree-level. \textbf{This is the reason why models that treat mesons at mean field obtain reasonable results.} However, the non-trivial mode is \(\Delta_1\), which corresponds to the true development of a Goldstone mode as a consequence of the breaking of the \(U(1)_{I_3}\) invariance.

\subsection{Parameter fixing and model setup}

The Linear Sigma Model with quarks contains several parameters that must be constrained by vacuum properties and symmetry considerations. A powerful constraint comes from the Ward-Takahashi identity relating the sigma and pion propagators,
\begin{equation}
D_{\sigma}^{-1} - D_{\pi}^{-1} = -2\lambda v^2,
\end{equation}
which, at tree level, implies the relation \(2g^2 = \lambda\) between the Yukawa and mesonic couplings. This identity also yields a mass relation in the vacuum,
\begin{equation}
m_\sigma^2 = 4m_f^2 + m_\pi^2,
\end{equation}
where \(m_f = g v\) is the constituent quark mass. These relations ensure consistency with chiral symmetry and reduce the number of independent parameters.

After imposing these constraints, the model retains only one free parameter, which we choose to be the vacuum quark mass $m_{0f}$. To ensure that our effective model reproduces the physical vacuum and finite-temperature behavior predicted by lattice QCD, we calibrate the remaining parameters against recent lattice results. The vacuum quark mass is varied between $200$–$300$ MeV to explore its impact on the phase structure and thermodynamic observables. Table~\ref{tab:parameters} summarizes the representative parameter sets obtained from this calibration procedure.

\begin{table}[htbp]
\centering
\caption{Model parameters used in this work, calibrated to reproduce lattice QCD results from recent studies.}
\begin{tabular}{lccc}
\hline
\textbf{Reference} & \(\lambda\) & \(m_\sigma\) (MeV) & \(m_f\) (MeV) \\
\hline
JHEP 07, 055 (2022) \cite{Brandt2023} & 8.33 & 423.792 & 200 \\
Phys. Rev. Lett. 134, 1 (2025) \cite{Abbott2025} & 18.74 & 616.117 & 300 \\
\hline
\end{tabular}
\label{tab:parameters}
\end{table}

With these parameters fixed, we explore a wide range of isospin chemical potentials, up to $\mu_I/m_\pi \sim 4$. This allows us to study the full phase structure, from the onset of pion condensation in the Bose-Einstein condensed regime up to a moderately high-density region where the equation of state starts approaching the BCS-like behavior expected from perturbative QCD.

\subsection{Speed of sound}

The squared speed of sound is a fundamental thermodynamic observable that characterizes the stiffness of the equation of state. It is defined as
\begin{equation}
c_s^2 = \frac{\partial P}{\partial \epsilon} = \frac{\partial P / \partial \mu_I}{\partial \epsilon / \partial \mu_I},
\end{equation}
where \(P = -V_{\text{eff}}\) is the pressure and \(\epsilon\) is the energy density. In the isospin-asymmetric medium, both quantities receive contributions from the tree-level potential and the one-loop corrections from fermions and bosons.

\section{Numerical results and discussion}

\begin{figure}[htbp]
\centering
\includegraphics[width=0.8\textwidth]{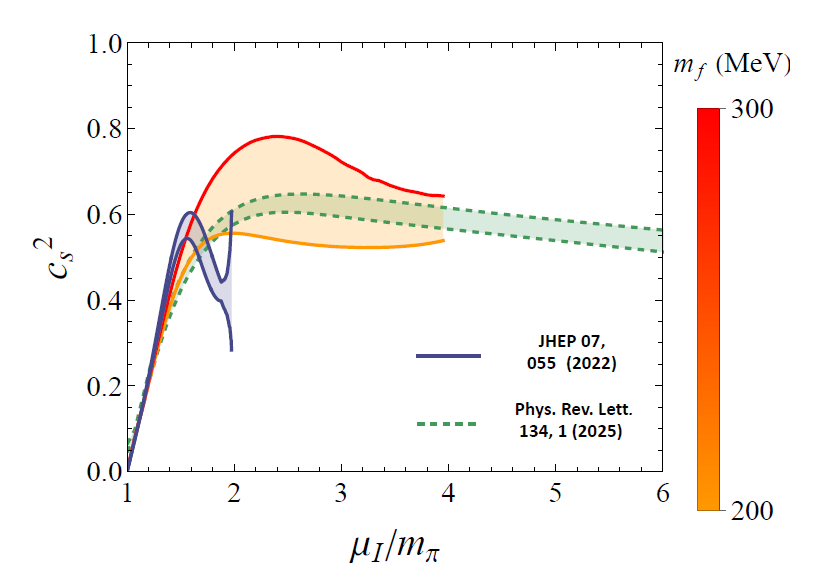}
\caption{Speed of sound squared \(c_s^2\) as a function of \(\mu_I/m_\pi\) for two different vacuum quark masses, compared with lattice QCD results from Abbott et al.~\cite{Abbott2023} and Brandt et al.~\cite{Brandt2023}. The peak structure and its dependence on \(m_f\) are clearly visible.}
\label{fig:cs2_comparison}
\end{figure}

We now present the numerical results for the speed of sound squared \(c_s^2\) as a function of the isospin chemical potential \(\mu_I\), obtained from the one-loop effective potential described in the previous sections. The calculations are performed using a continuous parameter scan for vacuum quark masses in the range \(m_f \in [200, 300]\) MeV, allowing us to explore the sensitivity of the results to this key parameter. Fig.~\ref{fig:cs2_comparison} shows a qualitative comparison of our results with LQCD simulations from recent studies~\cite{Brandt2023,Abbott2025}.

Several key observations emerge from Fig.~\ref{fig:cs2_comparison}:

\begin{itemize}
    \item The peak position and height obtained for intermediate values within the considered range in $m_f$ are in good agreement with the LQCD results of Ref.~\cite{Abbott2023}.
    
    \item The lower quark mass \(m_f = 200\) MeV shifts the peak to slightly lower values of \(\mu_I\) and reduces its magnitude. This brings the results closer to the lattice simulations of Ref.~\cite{Brandt2023}, and illustrates the sensitivity of the peak structure to the explicit breaking scale.
    
    \item Remarkably, both parameter sets capture the qualitative feature of a peak exceeding the conformal bound \(c_s^2 = 1/3\). This confirms that the enhancement of the speed of sound is a robust prediction of the one-loop LSMq and not an artifact of a particular parameter choice.

\end{itemize}

\section{Conclusions}

The results presented above demonstrate that the one-loop Linear Sigma Model with quarks successfully captures the non-perturbative physics of isospin-asymmetric matter. The pronounced peak in the speed of sound arises from the intricate interplay between the chiral condensate \(v\) and the pion condensate \(\Delta\), mediated by their mixing in the boson propagator. This mixing, encoded in the off-diagonal entries of the inverse propagator matrix, generates a nontrivial structure in the thermodynamic potential that is completely absent at tree level. It is precisely this quantum-induced mixing that drives the speed of sound above the conformal bound, bringing the model into quantitative agreement with lattice simulations.

The dependence on the vacuum quark mass \(m_f\) provides valuable insights into the role of explicit chiral symmetry breaking. By varying \(m_f\) while keeping other vacuum observables fixed, we effectively probe the sensitivity of the phase diagram to the strength of the chiral condensate and the corresponding curvature of the effective potential in the condensed phase. The resulting shift in the peak position in both its height and location in \(\mu_I\)  demonstrates that thermodynamic quantities in the condensed phase are sensitive to the vacuum parameters of the model, providing a way to calibrate effective models: by requiring that its predictions for observables such as the speed of sound match available lattice data, one can identify the range of \(m_f\) that yields the most consistent description.

For further reading on the subject of pion condensation in isospin-asymmetric matter, we refer the interested reader to the previous related works in Refs.~\cite{Ayala2024, Ayala2025, Avanicini2019, Lopes2025, Azeredo2026}. They provide complementary perspectives and extended discussions that deepen the understanding of this type of system from the point of view of effective models.

Looking ahead, the incorporation of additional degrees of freedom such as the \(\rho\) meson and other resonances will be essential for extending the validity of the model to higher isospin chemical potentials. These degrees of freedom are expected to become relevant once \(\mu_I \geq 4m_\pi\), where the characteristic energy scales approach the mass of the lightest vector mesons. Work in this direction is currently in progress and will be reported elsewhere.

\section*{Acknowledgments}

L.C.P acknowledges the financial support of a fellowship granted by Secretaría de Ciencia, Humanidades, Tecnología e Innovación (SECIHTI) as part of the Sistema Nacional de Posgrados. A.A. thanks the colleagues and staff of Universidade de São Paulo, of Instituto de F\'isica Te\'orica, UNESP and of Universidade Cidade de São Paulo for their kind hospitality during a sabbatical stay. A.A. also acknowledges support from the PASPA program of DGAPA-UNAM for the sabbatical stay during which this research was carried out. Support for this work has been received in part by a SECIHTI-M\'exico grant number CIORGANISMOS-2025-17 and by the DGAPA-PAPIIT-UNAM grant number IG100826. This work was partially supported by Conselho Nacional de Desenvolvimento Cient\'{\i}fico e Tecnol\'ogico (CNPq), Grants No. 312032/2023-4, 402963/2024-5 and 445182/2024-5 (R. L. S. F.), No. 141270/2023-3 and 201300/2025-7 (B. S. L.); 
Fundação de Amparo à Pesquisa do Estado do Rio Grande do Sul (FAPERGS), Grant No. 24/2551-0001285-0 (R. L. S. F.);
Instituto Nacional de Ci\^encia e Tecnologia---F\'isica Nuclear e Aplica\c{c}\~oes (INCT - FNA), Grants No. 464898/2014-5 and 408419/2024-5;
Ser\-ra\-pi\-lhei\-ra Institute (Grant No. Serra---2211-42230);
B. S. L. and R. L. S. F. also acknowledge the kind hospitality of the Center for Nuclear Research at Kent State University, where part of this work was done.

\end{document}